\documentclass[prb,twocolumn,showpacs,preprintnumbers,superscriptaddress]{revtex4-1}
\pdfoutput=1
\usepackage[usenames,dvipsnames]{color}
\usepackage{amssymb} 
\usepackage{times}
\usepackage{graphicx}
\usepackage{graphicx}
\usepackage{dcolumn}
\usepackage{datetime}
\usepackage{soul}
\usepackage{subfigure}
\usepackage[bookmarks, colorlinks=true, plainpages = false,citecolor = red, urlcolor = black, 
filecolor = blue]{hyperref}
\usepackage{multirow} 
\usepackage[english]{babel}
\usepackage{graphicx}
\usepackage{amsmath}
 
\usepackage[english]{babel}
\usepackage{graphicx}
\usepackage{amsmath}
 
\begin{document}
\title{Valley currents and non-local resistances of graphene nanostructures with broken 
inversion symmetry
from the perspective of scattering theory}
 
\author{George Kirczenow} 

\affiliation{Department of Physics, Simon Fraser
University, Burnaby, British Columbia, Canada V5A 1S6}

\date{\today}

\begin{abstract}\noindent

Phys. Rev. B {\bf 92}, 125425 Ð Published 17 September 2015. 

URL: http://journals.aps.org/prb/abstract/10.1103/PhysRevB.92.125425\\

Valley currents and non-local resistances of graphene nanostructures with broken inversion 
symmetry
are considered theoretically 
in the linear response regime. Scattering state wave functions of electrons entering 
the nanostructure from the contacts represented by groups of ideal leads are calculated by solving 
the Lippmann-Schwinger
equation and are projected onto the valley state subspaces to obtain the valley velocity fields 
and total valley
currents in the nanostructures. In the tunneling regime when the Fermi energy is in the spectral 
gap around the Dirac point energy, inversion symmetry breaking is found to result in 
strong enhancement of 
the non-local 4 terminal B\"{u}ttiker-Landauer resistance 
and in valley currents several times stronger than the conventional electric
current. 
These strong valley currents are the direct result of the injection of electrons from a contact
into the graphene in the tunneling regime. They are chiral and occur near contacts from 
which electrons are injected into the
nanostructure whether or not a net electric current flows through the contact. It is 
also pointed out that
enhanced non-local resistances in the linear response regime are not a signature of 
valley currents arising from 
the combined effect of the electric field and Berry curvature on the velocities of electrons.

\end{abstract}

 
\maketitle
\section{Introduction}

Graphene is a single atomic layer of carbon atoms arranged on a honeycomb lattice.
Since the early work of Wallace\cite{Wallace} it has been known that the electronic 
energy bands of graphene 
near the Fermi energy take the form of  a degenerate pair of cones, also referred to as `valleys.' 
The electronic density 
of states of graphene vanishes at the Dirac point energy which coincides with the Fermi 
energy in pristine graphene.
More recently it was recognized that the electronic structure of graphene has topological properties 
that relate to the Berry phase and Berry curvature.\cite{XiaoReview}
Specifically, graphene has a non-zero Berry phase associated with closed paths in 
reciprocal space that enclose a valley
apex.\cite{Ando} If the inversion symmetry of the graphene lattice is broken, for example, 
by a staggered 
potential at the two atoms of the unit cell, then the Berry curvature $\bf{\Omega}_{\bf{k}}$ 
becomes non-vanishing 
near the apex of each valley.\cite{XiaoPRL} In semiclassical theories
of electron transport in graphene, the electron velocity ${\bf v}_{\bf k}$ is related to the 
Berry curvature by\cite{KL,XiaoReview,CN1,CN2,Chong}
\begin{equation}\label{vomega}
 {\bf v}_{\bf k}=\frac{1}{\hbar} \frac{\partial \epsilon_{{\bf k}}}{\partial {\bf k}} 
 +\dot{{\bf k}} \times \bf{\Omega}_{\bf{k}}
\end{equation}
where $\epsilon_{{\bf k}}$ is the energy
of a Bloch state with wave vector ${\bf k}$ and, in the absence of magnetic fields, 
\begin{equation}\label{Efield}
 \hbar \dot{{\bf k}} = q_e {\bf E}
\end{equation}
where $q_e$ is the electron charge and 
${\bf E}$ is the electric field. Since $\bf{\Omega}_{\bf{k}}$ points in opposite 
directions in the two valleys, Eqs. \ref{vomega} and \ref{Efield} imply that, 
in the presence of an electric field ${\bf E}$, electrons in the two valleys of 
graphene with broken inversion symmetry  will have
differing velocities. This difference in velocity might, in principle, 
be used to separate electrons belonging to the different valleys spatially, 
and thus be employed in future valleytronic 
devices in which the valley index of electrons plays a role 
somewhat analogous to that of the electron spin 
quantum number in spintronic devices.  

Recently, Gorbachev {\em et al.}\cite{Gorbachev} carried out non-local 4-terminal 
resistance measurements on graphene placed on a hexagonal boron nitride
(hBN) substrate, varying the graphene Fermi level by the application of a gate voltage. 
They found a striking 
enhancement of the measured non-local resistance when the Fermi level passed through Dirac 
points for samples with the 
crystallographic axes of the graphene aligned with those of the hBN. They interpreted the 
enhanced non-local 
resistance as a signature of valley currents in their samples based on the following intuitive 
picture:\cite{Gorbachev} 
The aligned hBN substrate breaks the inversion symmetry of 
graphene.\cite{Giovannetti,Sachs,Kindermann,Song} 
This results in non-zero Berry curvatures for states close 
in energy to Dirac points. Then, according to Eqs. \ref{vomega} and \ref{Efield}, 
an electric field ${\bf E}$
that drives electric current through the sample would result in differing electron 
velocities in the two graphene 
valleys. This would imply a net valley current transverse to the electric current 
when the Fermi level is close to a Dirac point energy.\cite{Gorbachev} It was further 
argued\cite{Gorbachev} that 
this valley current flows into the region of the sample 
between the voltage probes and induces an electric field there,
resulting in a potential difference between the voltage probes and in an
enhancement of the non-local resistance, consistent with the experiment.\cite{Gorbachev} 

However, valley currents were not observed {\em directly}
in the experiment of Gorbachev {\em et al.}\cite{Gorbachev}
Also, the interaction between
the hBN and graphene is expected theoretically\cite{Giovannetti,Sachs,Kindermann,Song} 
to open an energy gap
in the electronic density of states of graphene around the Dirac point energy, where the 
Fermi level is located
when the observed\cite{Gorbachev} enhanced non-local resistance has its maximum value. 
The presence of this gap is supported
by the observation of activated transport by Gorbachev {\em et al.}\cite{Gorbachev} in 
some of their samples. 
Within the gap the transport mechanism is quantum tunneling, a phenomenon that has no classical 
analog. Therefore
the applicability of the semiclassical Eqs. \ref{vomega} and \ref{Efield} to this regime is unclear  
and a fully quantum mechanical theory of transport is necessary.

A rigorous, fully quantum mechanical approach to calculations of 
multi-terminal transport coefficients
of nanostructures is provided by B\"{u}ttiker-Landauer theory.\cite{BL} However,  
the conceptual framework of B\"{u}ttiker-Landauer theory differs from that 
of the topological arguments\cite{Gorbachev} that have been outlined above.
For example, within B\"{u}ttiker-Landauer 
theory, in the linear response regime (i.e., in the limit where the applied bias voltages and currents
approach zero), the four-terminal resistances depend only on the 
set of quantum electron transmission probabilities $\{T_{i,j}\}$ between all pairs of contacts 
$\{i,j\}$ and the reflection probabilities $\{R_{i,i}\}$ at contacts $\{i\}$ evaluated at the
Fermi energy (or, at finite temperatures, at energies near the Fermi energy). 
In this limit, the electric field ${\bf E}$ that appears in Eq. \ref{Efield} goes to zero,
and therefore in this limit it has no effect on $\{T_{i,j}\}$ and $\{R_{i,i}\}$. Therefore, according to 
B\"{u}ttiker-Landauer theory, the topological mechanism of valley currents 
that is embodied in the term $\dot{{\bf k}} \times \bf{\Omega}_{\bf{k}}$ in Eq.  
\ref{vomega} has {\em no} effect on four-terminal resistances 
(or more generally on two-, three- or
other multi-terminal resistances, including non-local resistances) in the linear 
response regime. Thus it follows from B\"{u}ttiker-Landauer 
theory that non-local resistance measurements
in the linear response regime can not provide experimental evidence of 
topological effects arising from the electric field in Eq. \ref{Efield}. 

In other words, B\"{u}ttiker-Landauer theory shows that
non-local (and other) resistances measured in the linear response regime do not depend on 
whether electrons travel through the sample under the influence of an electric
field due to applied bias voltages or simply scatter through the sample freely
in the energy window between the highest and lowest contact electrochemical 
potentials without being subjected to any driving electric field. This means that the
effects of the topological term $\dot{{\bf k}} \times \bf{\Omega}_{\bf{k}}$ in Eq.  
\ref{vomega} that arise from the driving electric field cannot be detected by 
local or non-local resistance measurements in the linear response
regime. 

It is therefore of interest to explore theoretically the multi-terminal resistances and valley currents
in fully quantum mechanical models of graphene devices with broken inversion symmetry from
the perspective of B\"{u}ttiker-Landauer theory. This is done in the present 
paper for transport in graphene
nanostructures subjected to staggered potentials in the linear response regime. 
It will be shown here that
the application of a staggered potential results in strong enhancement of the non-local
four-terminal resistance when the Fermi level is close to the Dirac point energy,
in qualitative agreement with the experimental findings of Gorbachev {\em et al.}\cite{Gorbachev}
Despite the studied nanostructures having atomically abrupt boundaries where 
electrons scatter strongly and crystal momentum is not conserved, valley currents, up to 
several times larger than the conventional electric current, will be shown to appear 
in response to electrochemical potential differences between the electrodes
when the Fermi level is  
near the Dirac point  of  
graphene nanostructures with broken inversion symmetry. 
These large valley currents are not generated by the topological
mechanism embodied in Eqs. \ref{vomega} and \ref{Efield} since the present 
calculations are in the linear response limit where the driving electric field 
${\bf E}$ in Eq. \ref{Efield} tends to zero
and therefore it does not appear in the Hamiltonian of the system in these 
B\"{u}ttiker-Landauer 
theory-based calculations. Because these strong valley currents occur in a gap 
in the energy
spectrum of the nanostructure they require electron tunneling and consequently their strength
decays rapidly as the distance from a contact that injects electrons into the graphene
nanostructure increases. These valley currents will be seen to be chiral and to travel 
along the edge of the graphene that is in contact with an electrode 
that injects electrons into the graphene. If electrons are injected into the graphene nanostructure
with broken inversion symmetry from a scanning tunneling microscope (STM) tip 
the valley currents will be shown to 
form a vortex circulating around the
location at which the electron injection occurs.
At Fermi energies further from the Dirac point and outside of the gap in the density of states
of the nanostructure, valley currents are also induced by bias voltages applied to the
nanostructure. However in this regime they are found to be weaker, not to require
inversion symmetry breaking, and to extend into regions of the nanostructure
that are not close to the contacts. 

The remainder of this paper is organized as follows. In Section \ref{MF} 
The model of graphene nanostructures with broken inversion
symmetry coupled to current and voltage contacts that is studied in this work is presented. 
B\"{u}ttiker-Landauer theory and the Lippmann-Schwinger equation and how they apply to 
this model are outlined. Valley currents, valley velocity fields and non-local resistances are defined
and the methodology used to calculate them is described. The numerical results 
obtained from these calculations are
presented in Section \ref{Results}. The significance of the present findings 
is discussed in Section \ref{Discussion}.
  
\section{Model and Formalism} 
\label{MF}
 In this paper the graphene nanostructures will be described by the 
 standard nearest neighbor 
 tight-binding Hamiltonian on a 
 honeycomb lattice,
\begin{equation}
 H_{\text{GN}}=\sum_{n}\epsilon _{n}a_{n}^{\dag }a_{n}-\sum_{\left\langle n,m\right\rangle }
 t_{nm}\left( a_{n}^{\dag }a_{m}+h.c. \right),
 \label{eq:hamiltonian}
\end{equation}%
where $\epsilon _{n}$ is the on-site energy, $t_{nm}=t=2.7$ eV defines the matrix 
element between $p_z$ orbitals on nearest-neighbor atoms and the spin index is suppressed. 
This Hamiltonian with $\epsilon _{n}=0$ is known to describe the $\pi$ band dispersion of 
graphene well at low energies,\cite{Reich02}
and has been used in numerous studies of electron transport in graphene 
nanostructures.\cite{review}
In order to introduce inversion symmetry breaking into the model, the simple 
choice $\epsilon _{n}=\pm \Delta$ is made
so that $\epsilon _{n}$ is positive on one atom of the graphene unit cell and negative on the 
other. The amplitude of the symmetry
breaking energy is chosen to be $\Delta=0.0602$ eV consistent with estimates for 
graphene on hBN reported 
in Ref. \onlinecite{Sachs}.

This idealized model has been chosen since the purpose of this paper is to investigate the 
fundamental effects 
of inversion symmetry breaking in its simplest form on multi-terminal transport coefficients 
and valley currents within 
the fully quantum mechanical B\"{u}ttiker-Landauer framework. 
It should be noted that the graphene on hBN system is more involved since the lattice 
parameters of hBN and graphene have a 2\% mismatch and local configurations
with N atoms under the centers of graphene hexagons and B atoms under C atoms have 
the lowest energy.\cite{Sachs} However,
these complications will not be considered here.

In B\"{u}ttiker-Landauer theory,\cite{BL} at zero temperature and in the linear response regime,
the currents $I_i$ flowing towards the nanostructure in contacts $i$ are related to the 
electrochemical potentials $\mu_i$ of the contacts by
\begin{equation}
 I_i=\frac{q_e}{h}(N_i\mu_i-\sum_j T_{i,j}\mu_j)
 \label{eq:buttiker}
\end{equation}%
where $N_i$ is the number of electronic modes incident on the nanostructure from 
contact $i$ and $T_{i,j}$ is the multichannel electron transmission probability from 
contact $j$ to contact $i$. 
$T_{i,i} =R_{i,i}$ is the multichannel electron reflection probability from the 
nanostructure in contact $i$. 

\begin{figure}[t!]
\centering
\includegraphics[width=1.0\linewidth]{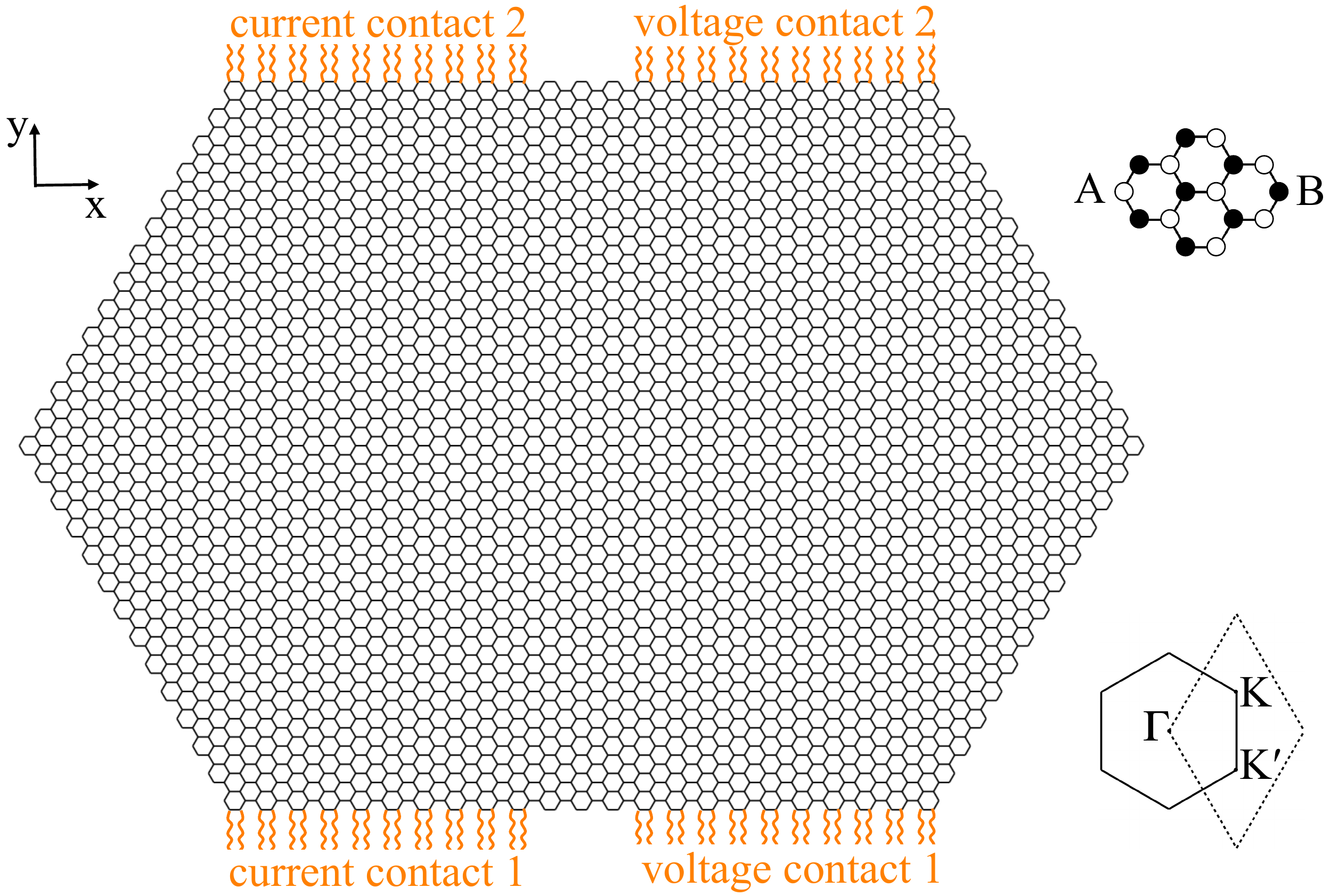}
\caption{
(Color online) Graphene nanostructure with armchair edges. 
The size of the nanostructure in the $y$-direction
is 9.838 nm. Electron stream is injected
through current contact 1 and exits through current contact 2. There is no net
electric current entering or leaving through the voltage contacts 1 and 2; the potential difference
between them is measured. Wavy lines represent ideal semi-infinite 1D 
leads connecting graphene C
atoms to electron reservoirs. Upper right inset: The graphene sublattices A (B) = open (filled)
circles. Lower right inset: Hexagonal (solid) and rhombic (dotted) Brillouin zones of 
graphene. K and K$'$
are the two Dirac points. 
}
\label{nanostructure} 
\end{figure}

The coefficients $T_{i,j}$ that enter the B\"{u}ttiker-Landauer theory  
are calculated in this paper as in many previous theoretical studies of quantum 
transport in nanostructures
\cite{Buker,Firuz1,Firuz2,Firuz3,Cardamone,Piva1,Piva2,Dalgleish1,
George2,Renani,Saffar,AFG,Majumder,Geo}
with semiconducting,\cite{Piva1,Piva2,
George2,Majumder,Geo}
molecular,\cite{Buker,Firuz1,Firuz2,Firuz3,Cardamone,Piva1,Piva2,Dalgleish1,
George2,Renani,Saffar,AFG,Geo} metallic,
\cite{Buker,Firuz1,Firuz2,Firuz3,Cardamone,Piva1,Piva2,Dalgleish1,
George2,Renani,Saffar,AFG,Majumder,Geo} magnetic,
\cite{Renani,Saffar,Majumder,Geo} and carbon-based \cite{Saffar}
constituents. 
Each contact is represented by a set of ideal semi-infinite one-dimensional (1D) tight-binding leads 
(the wavy lines in Fig. \ref{nanostructure}) with one
orbital per site and nearest neighbor hopping. One such ideal lead is 
attached to each peripheral site of the graphene nanostructure that is adjacent to a contact. 
The Hamiltonian of lead $n$ is
\begin{equation}
 H_{\text{L}n}=\sum_{r}\epsilon _{n}b_{r}^{\dag }b_{r}
 -\sum_{\left\langle r,s\right\rangle }t\left( b_{r}^{\dag }b_{s}+h.c. \right),
 \label{eq:lead}
\end{equation}%
where $t$ is the same as in $H_{\text{GN}}$ (Eq.  \ref{eq:hamiltonian}). 
The site energy $\epsilon _{n}$ in Eq.  \ref{eq:lead}
is the same as that of the site of the graphene nanostructure to which the lead is connected. 
The coupling Hamiltonian between lead $n$ and the edge site of the graphene nanostructure is
\begin{equation}
 W_{n}=-t ( b_{n}^{\dag }a_{n}+h.c. ),
 \label{eq:couple}
\end{equation}%
The quantum transmission amplitudes for an electron to scatter at energy $E$
via the nanostructure from one 1D ideal lead to another are found
by solving the Lippmann-Schwinger equation
\begin{equation}
|\psi^m\rangle=|\phi_0^m\rangle+G_0(E)W|\psi^m\rangle\  ,
\label{LS}
\end{equation}
where $|\phi_0^m\rangle$ is an electron eigenstate of the $m^\text{th}$
ideal semi-infinite lead that is decoupled
from the graphene nanostructure, $G_0(E)$ is the Green's function of
the decoupled system of the ideal leads and the graphene nanostructure,
and $W=\sum_{n} W_{n}$ is the coupling between the graphene nanostructure and the ideal
leads. $|\psi^m\rangle$ is the scattering eigenstate of the
complete coupled system associated with the incident electron
state $|\phi_0^m\rangle$. Then
\begin{equation}
T_{i,j}(E)=\sum_{n,m}|t^{ij}_{nm}(E)|^2 {v^i_n}/{v^j_m}
\label{T}
\end{equation}
where $t^{ij}_{nm}(E)$ is the quantum transmission amplitude [obtained from the scattering
state $|\psi^m\rangle$ defined by Eq. (\ref {LS})] for an electron at energy $E$
to scatter via the graphene nanostructure from ideal 1D lead $m$ of contact $j$ to ideal 
1D lead $n$ of contact $i$.
The sum is over ideal leads $n$ ($m$) in contact $i$ ($j$). ${v^{i(j)}_{n(m)}}$ is
the electron velocity in ideal 1D lead $n$ ($m$) of contact $i$ ($j$) at energy $E$; 
$v^i_n = \frac{1}{\hbar} \frac{\partial\epsilon}{\partial k} $ where $\epsilon$ are the energy 
eigenvalues of the Hamiltonian $H_{\text{L}n}$ [Eq. \ref{eq:lead}] of an infinite ideal 
1D tight binding chain.

Having evaluated $T_{i,j}$ at the Fermi energy in this way, the B\"{u}ttiker equations 
 \ref{eq:buttiker} are solved in the linear response regime to find the non-local 4-terminal
 resistance 
 \begin{equation}
 R_{\text{NL}} = \Delta V/I
 \label{R}
\end{equation}
 where $I$ is the current passing through the current contacts
 and $\Delta V = \Delta \mu/q_e$ is the potential difference between the voltage contacts;
 see the contacts in Fig. \ref{nanostructure}.
 
 The valley currents induced in the nanostructure in response to to the electrochemical potential
 differences between the various contacts in the linear response regime are estimated 
 as follows: The scattering state $|\psi^m\rangle$ of electrons injected into the 
 nanostructure from ideal 1D lead $m$ is calculated at the Fermi energy for every
 lead $m$ by solving the Lippmann-Schwinger equation (\ref{LS}).  Then the scattering
 state $|\psi^m\rangle$ is projected onto the subspaces of 
 Bloch states of graphene that belong to the
 K and K$'$ valleys. This yields the projected valley states, $|\psi^m_\text{K}\rangle$ and
 $|\psi^m_{\text{K}'}\rangle$, respectively. For this purpose a Bloch state is assigned 
 to the K (K$'$) valley 
 if its wave vector lies within the upper (lower) half of the rhombic Brillouin zone
 defined by the dotted boundary in the lower right inset of Fig. \ref{nanostructure}. 
 
 The $\xi$-component of the velocity operator for electrons within the graphene nanostructure is
 \begin{equation}
 {v_\xi} =\frac{1}{i\hbar}[\xi,H_{\text{GN}}]
 \label{v}
\end{equation}
where $ \xi =\sum_{p} \xi _{p} a_{p}^{\dag }a_{p}$
and $\xi _{p}$ is the $\xi-$coordinate of atomic site $p$ of the graphene nanostructure.
The expectation value of ${v_\xi}$ in the graphene nanostructure in the state
 $|\psi^m\rangle$ is then
 \begin{equation}
 \langle\psi^m|{v_\xi}|\psi^m\rangle =\frac{it}{2\hbar}
 \sum_{k,l}(\xi_k-\xi_l)(\psi^{m*}_{k}\psi^{m}_{l}-\psi^{m}_{k}\psi^{m*}_{l})
 \label{vav}
\end{equation}
where $k$ and $l$ are nearest neighbor sites of the graphene nanostructure, 
$\xi_l = x_l~\text{or}~y_l$,
$\psi^{m}_l= \langle Z_l |\psi^m \rangle$ and $|Z_l \rangle$
is the $2p_z$ orbital of the carbon atom at site $l$.

For a graphene nanostructure with multiple contacts $i$ each at its own electrochemical
potential $\mu_i$ the electron transport through the device is governed by a weighted 
average over the velocities
associated with the scattering states injected by the various ideal leads $m_i$ 
that make up all of the contacts $i$.
The relevant weighted average will be defined here as
 \begin{equation}
v_{\xi }=\sum_{m,i}\langle\psi^{m_i}|{v_\xi}|\psi^{m_i}\rangle {\Delta \mu_i} /\sum_{m,i}{\Delta \mu_i} 
 \label{weightedsum}
\end{equation}
where $\Delta \mu_i = \mu_i - \mu_\text{min}$ and $\mu_\text{min}$ is the 
electrochemical potential of the contact
with the lowest electrochemical potential.  

The weighted valley velocities $v_{\text{K}\xi }$ and $v_{\text{K}'\xi }$ for electrons 
in valleys K and K$'$ are defined similarly by replacing $|\psi^m \rangle$ and 
$|\psi^{m_i}\rangle$ in Eqs. (\ref{vav}) and (\ref{weightedsum}) by
their projections $|\psi^m_\text{K}\rangle$ and $|\psi^{m_i}_\text{K}\rangle$, and 
 $|\psi^m_{\text{K}'}\rangle$ and $|\psi^{m_i}_{\text{K}'}\rangle$ onto the valleys K and K$'$, 
 respectively. The weighted valley velocity is then
 defined as
 \begin{equation}
v^\text{val}_{\xi}=(v_{\text{K}\xi }-v_{\text{K}'\xi })
 \label{valleyvel}
\end{equation}

Equation  (\ref{vav}) expresses the expectation value of the electron velocity 
as a sum of terms evaluated
at pairs $(k,l)$ of nearest neighbor atoms of the graphene nanostructure. 
This suggests that each such
term be interpreted as the value of the velocity field for the scattering state 
$|\psi^m\rangle$ at the location
of each nearest neighbor pair. Assigning this value of the velocity field 
$v_\xi^\text{F}$ arbitrarily to the mid point 
$(x,y)=({(x_k+x_l)}/{2},{(y_k+y_l)}/{2})$ of the atomic pair yields
 \begin{equation}
 \langle\psi^m|{v_\xi^\text{F}}(x,y)|\psi^m\rangle =\frac{it}{\hbar}
 (\xi_k-\xi_l)(\psi^{m*}_{k}\psi^{m}_{l}-\psi^{m}_{k}\psi^{m*}_{l})
 \label{vavloc}
\end{equation}
The corresponding velocity field weighted by the contributions of the states contributing
to transport through the nanostructure is then, as in Eq.  (\ref{weightedsum}), given by
 \begin{equation}
v^\text{F}_{\xi }(x,y)=\sum_{m,i}\langle\psi^{m_i}|{v_\xi^\text{F}}(x,y)|\psi^{m_i}\rangle 
{\Delta \mu_i}/\sum_{m,i}{\Delta \mu_i} 
 \label{weightedsumfield}
\end{equation}
Obtaining the weighted velocity fields $v^\text{F}_{\text{K}\xi }$ and $v^\text{F}_{\text{K}'\xi }$ 
for the valleys K and K$'$ similarly by replacing $|\psi^m \rangle$ by
the projections $|\psi^m_\text{K}\rangle$ and
 $|\psi^m_{\text{K}'}\rangle$, the valley velocity field is defined as
 \begin{equation}
v^\text{val~F}_{\xi }(x,y)=v^\text{F}_{\text{K}\xi }(x,y)-v^\text{F}_{\text{K}'\xi }(x,y)
 \label{valleyfield}
\end{equation}

 \begin{figure}[t!]
\centering
\includegraphics[width=1.0\linewidth]{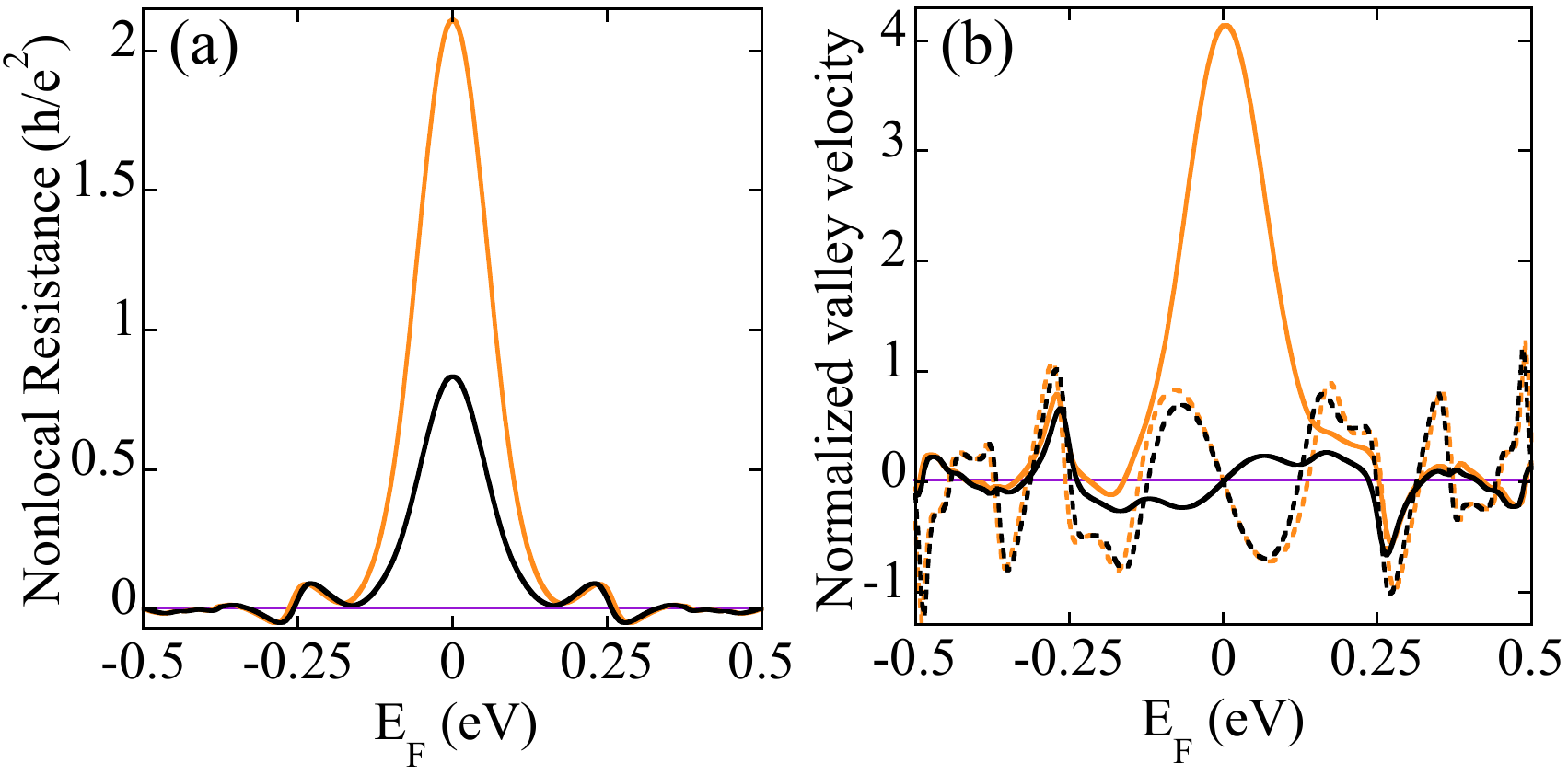}
\caption{
(Color online) Calculated linear response properties vs. 
Fermi energy of the structure in Fig. \ref{nanostructure}
at zero temperature. 
Current $I$ flows through the current contacts with no net current through
either voltage contact. Results for symmetry breaking parameter 
$\Delta=0.0602$ (0.0) eV  are orange (black). 
(a) Nonlocal resistance $R_{\text{NL}}$ [Eq.
 (\ref{R})]. (b) Normalized valley velocities $v^\text{val}_{x}/v_{y}$ and 
 $v^\text{val}_{y}/v_{y}$ [Eq. (\ref{weightedsum}),
 (\ref{valleyvel})] are 
 solid and dashed lines, respectively.
}
\label{RNLandValcur} 
\end{figure}

\section{Results} 
\label{Results}
The results of B\"{u}ttiker-Landauer calculations of the non-local four-terminal
resistance $R_{\text{NL}}$ defined by Eq. (\ref{R}) for the structure in Fig. \ref{nanostructure}
at zero temperature in the linear response regime are
shown in Fig. \ref{RNLandValcur}(a) as a function of the Fermi energy $E_\text{F}$.
The Fermi level crosses the Dirac point energy at $E_\text{F} =0$.
Near the Dirac point,  $R_{\text{NL}}$ for the model with $\Delta=0.0602$eV (
orange line) exceeds $R_{\text{NL}}$ for $\Delta=0$ (black line)
by a factor of $\sim2.5$. 
Thus within B\"{u}ttiker-Landauer theory, the breaking of inversion
symmetry of the graphene unit cell results in strong enhancement of the non-local
resistance near the Dirac point. However, as can also be seen in Fig. \ref{RNLandValcur}(a),
well away from the Dirac point energy the inversion
symmetry breaking has little effect on the non-local
resistance.

 \begin{figure*}[t]
\centering
\includegraphics[width=1.0\linewidth]{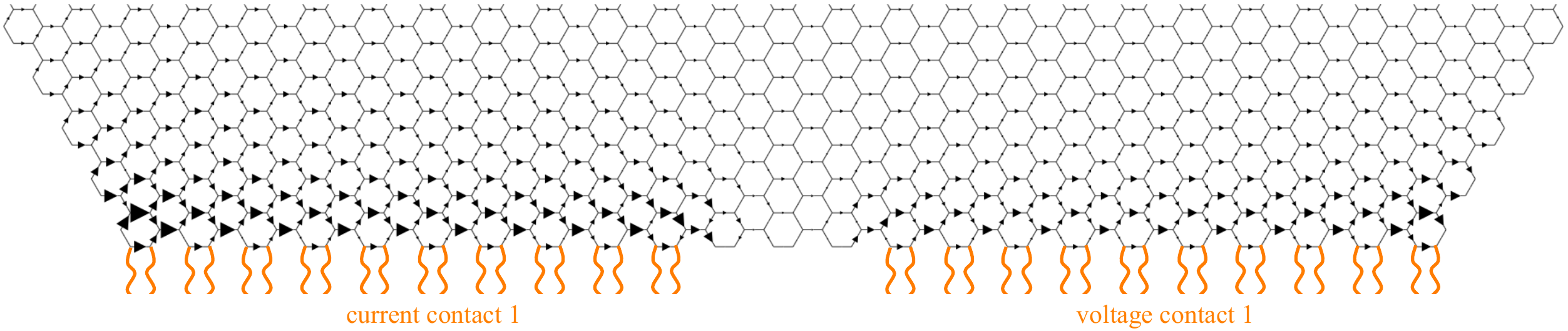}
\caption{
(Color online) Valley velocity field [Eq.  (\ref{valleyfield})] in the lower part of the 
structure in Fig. \ref{nanostructure} for electron flow from current contact 1 to current 
contact 2 (shown in Fig. \ref{nanostructure}). For electron flow in this direction, the valley velocity
field is much stronger in the vicinity of the contacts shown here 
than elsewhere in the graphene 
nanostructure. $\Delta=0.0602$eV. $E_\text{F}=0$.}
\label{valvelfield} 
\end{figure*}

Since the net electron flow in the structure in Fig. \ref{nanostructure} is from current contact 1
to current contact 2 (i.e., in the $y$-direction), the weighted velocity vector
$\vec{v}$ (Eq. \ref{weightedsum}) is expected to point in the $y$-direction. Accordingly, $v_x$
in Eq. \ref{weightedsum} is found to be zero within numerical error in the present computations.
The computed normalized, weighted valley velocities 
$v^\text{val}_{x}/v_y ~\text{and}~ v^\text{val}_{y}/v_y$ are
shown in Fig. \ref{RNLandValcur}(b). They are found to be non-zero except 
at isolated values of the Fermi energy both in the presence and absence of 
symmetry breaking. Note that $v^\text{val}_{\xi}/v_y = I^\text{val}_{\xi}/I$
where $I^\text{val}_{\xi}$ is the ${\xi}$-component of the valley current and $I$ is 
the total electric current through the nanostructure.
The most striking feature of Fig. \ref{RNLandValcur}(b) is the strong 
peak near the Dirac point ($E_\text{F} =0$) of 
 $v^\text{val}_{x}/v_y = I^\text{val}_{x}/I$ for broken inversion symmetry, (the solid orange curve). 
 At its maximum
 the valley current in the $x$-direction exceeds the total conventional electric current through 
 the nanostructure
 by a factor of more than 4. This peak of the valley current is in the 
 gap (of width 0.264 eV) in the 
 energy spectrum of the broken symmetry nanostructure around the Dirac point.
  By contrast, the $x$-component of the valley current in the absence of
 symmetry breaking (solid black curve) and the $y$-component of the valley 
 current with (orange dashed curve) 
 and without (black dashed curve) symmetry breaking all vanish at the Dirac point and 
 are relatively weak
 elsewhere. As can also be seen in Fig. \ref{RNLandValcur}(b), 
 for $E_\text{F}$ well away from the Dirac point energy, both 
 $v^\text{val}_{x}/v_y ~\text{and}~ v^\text{val}_{y}/v_y$
 are insensitive to the breaking of the inversion symmetry of the graphene.
 
  The full widths of the central peaks of both the non-local resistance 
 and the normalized valley velocity $v^\text{val}_{x}/v_{y}$ for the system with 
 broken inversion symmetry [the orange curves in Figs. \ref{RNLandValcur}(a) and (b),
 respectively] are close in size to the 0.264eV gap in the energy spectrum of 
 the broken symmetry nanostructure. Because the value of the symmetry breaking parameter
 is relatively small ($\Delta=0.0602$eV), the size of the spectral gap is determined mainly 
 by the quantum confinement of the electrons in the graphene nanostructure and the 
 armchair character of the nanostructure's edges.\cite{review} Thus for the same nanostructure
 but with the symmetry breaking turned off ($\Delta=0$) the width of the energy gap has a
similar value, 0.235 eV. For this reason the main nonlocal resistance peak 
in Fig. \ref{RNLandValcur}(a)
has almost the same width for $\Delta=0$ (the black curve) as for $\Delta=0.0602$eV
(the orange curve).   
   
 The valley velocity field $\vec{v}^\text{~val~F}(x,y)$ is shown in Fig. \ref{valvelfield} 
 for the lower part of the nanostructure 
 in Fig. \ref{nanostructure}. Inversion symmetry is broken and the Fermi level is 
 at the Dirac point. The
 valley velocity is large near current contact 1 and voltage contact 1. 
 Its magnitude initially increases 
 but then decreases rapidly with 
 increasing distance from the contacts. The valley velocity is clearly chiral, 
 pointing mainly from left to right 
 near the graphene boundary shown (its overall direction reverses if the sign of $\Delta$, 
 the symmetry breaking parameter,
 is changed) but it does not extend along the boundary much beyond where a contact ends.
  
 In Fig. \ref{valvelfield} the electron flow enters the graphene 
 nanostructure through current contact 1
 and exits through current contact 2 that is located outside of 
 the region shown in Fig. \ref{valvelfield};
 see Fig. \ref{nanostructure} for its location. If the direction of the electron 
 flow through the graphene nanostructure
 is reversed, so that electrons flow instead from current contact 2 to current 
 contact 1 (in Fig. \ref{nanostructure}),
 then the valley velocity field becomes strongest near current contact 2 and 
 voltage contact 2, i.e., 
 near the opposite edge of the sample to that where the valley velocity field 
 is strongest in Fig. \ref{valvelfield}.
 For electron flow from current contact 2 to current contact 1 the direction of 
 the valley velocity field
 is from right to left, i.e., its direction is opposite to that in Fig. \ref{valvelfield}, 
 consistent with
 the chiral character of the valley current.
 
\begin{figure}[t!]
\centering
\includegraphics[width=1.0\linewidth]{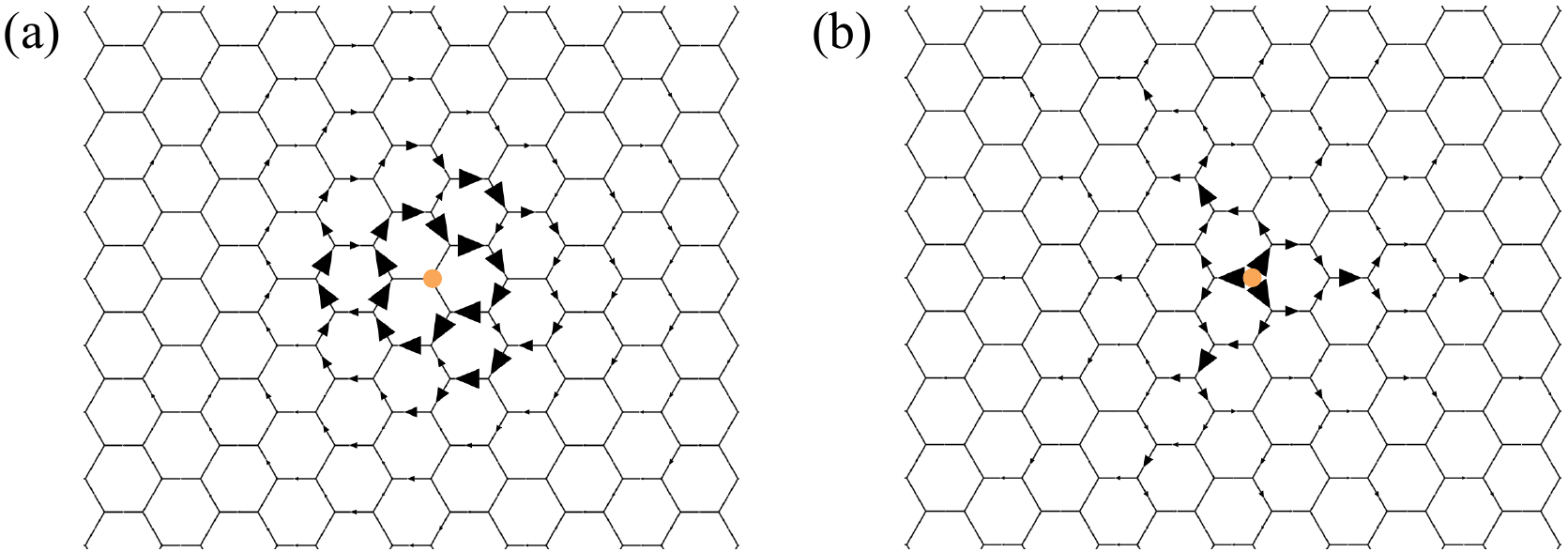}
\caption{
(Color online) (a) Valley velocity field [Eq.  (\ref{valleyfield})] and (b) velocity field 
[Eq.  (\ref{weightedsumfield})]
for electrons injected into a graphene nanostructure with broken inversion 
symmetry via a single carbon atom (orange). 
Only a small part of the graphene nanostructure is shown.
$\Delta=0.0602$eV. $E_\text{F}=0$.
}
\label{STM} 
\end{figure}

 The chiral nature of the valley current is further clarified in Fig. \ref{STM}. 
 Fig. \ref{STM} 
 (a) shows the strongest part of the valley
 velocity field associated with electron injection into a graphene nanostructure with broken 
 inversion symmetry 
 via a single interior carbon atom (colored orange) of the nanostructure, as in an idealized STM 
 setup. The valley velocity
 field forms a vortex circulating clockwise (counter-clockwise if the sign of $\Delta$ is changed)
 around the injection point, whereas the electron flux travels outwards overall
 from the injection point as shown in Fig. \ref{STM} (b).
 
 In the vicinity of 
 a contact the valley current is due almost entirely to electrons injected into the 
 nanostructure from that
 contact. This is true even for the voltage contact (Fig. \ref{valvelfield}) 
 through which no $net$ electric 
 current flows since the zero net
 current is due to equal fluxes of electrons entering and leaving the contact. 
 The valley current of electrons
 leaving the contact is much larger than that of those entering the contact. 
 Thus a contact through which
 no net electric current flows can be used to create a valley current into a graphene 
 nanostructure with 
 broken inversion symmetry. I.e., it can in principle generate a pure valley current.
 
The results presented above have been for a graphene nanostructure with only armchair 
edges. Similar calculations
for a rectangular structure of similar
 size (dimensions
 9.656nm $\times$ 9.838nm) and similar contacts but with two zigzag 
 and two armchair edges were carried out and yielded qualitatively 
 similar results but the effects of inversion symmetry breaking were 
 found to be much stronger in this case: The non-local resistance 
 for  $\Delta=0.0602$eV was 
 found to exceed that for $\Delta=0$
by more that a factor of 100 for $E_\text{F}$ near the Dirac point energy. 
Also, at its maximum the valley current 
in the $x$-direction was found to exceed the total conventional electric
current through the nanostructure by a factor of more than 18. These large numbers
are attributable to the flat electronic dispersion at graphene zigzag edges\cite{flatband} 
in tight binding models
described by the non-interacting electron Hamiltonian, Eq. \ref{eq:hamiltonian}. However,
theoretical studies have suggested that electron-electron interactions 
may give rise to magnetism at zigzag edges.\cite{magnetism,Lado} 
The potential implications of this for valley currents and nonlocal 
resistances are beyond the scope of the present paper.

Different approaches for realizing valley currents in graphene have 
also been proposed based on
electric fields acting on electrons in the presence of Berry curvature,\cite{XiaoPRL} 
graphene point contacts with zigzag edges,\cite{Rycerz} electron scattering at the 
boundary between monolayer and bilayer graphene,\cite{Nakanishi}
electron scattering at a line defect in graphene,\cite{Gunlycke} illumination of
monolayer\cite{Oka} or bilayer\cite{Abergel} graphene by circularly polarized radiation,
optical injection of a pure valley current in graphene,\cite{Golub} and gate-induced 
valley filtering in bilayer graphene.\cite{Costa} Strong valley current polarizations, 
comparable to those obtained with the present approach for the structure
having both zigzag and armchair edges have been estimated
for some of these approaches,\cite{Rycerz,Gunlycke,Abergel,Golub,Costa} 
albeit for models of infinite two-dimensional
graphene or infinite graphene ribbons. Also, unlike in the present
work, the spatial distribution of valley currents was not reported
in the previous studies,\cite{XiaoPRL, Rycerz, Nakanishi, Gunlycke, Oka, 
Abergel,Costa} and the effects of boundaries between the graphene
and source and drain electrodes were not taken into account. 
  
 \section{Conclusions} 
\label{Discussion}

The present work suggests an approach
for creating valley currents in graphene with broken inversion symmetry 
that differs fundamentally from previous 
proposals.\cite{KL,XiaoReview,CN1,CN2,Chong,Gorbachev,XiaoPRL,Rycerz,Nakanishi,
Gunlycke,Oka,Abergel,Golub,Costa,YXN} 
As has been explained above, it follows from B\"{u}ttiker-Landauer 
theory that in the linear response regime considered here,
the transport properties of nanostructures are determined by electron scattering states
calculated in the limit where the driving electric field has been sent to zero. 
Consequently in the linear 
response regime the
acceleration of electrons by the driving electric field has no effect on multi-terminal resistance
coefficients or on $\vec{I}^\text{val}/I$, the ratio of the valley current
$\vec{I}^\text{val}$ and the conventional electric current $I$ passing through the nanostructure.
Thus
the valley currents discussed here are not due to electron acceleration in an electric
field in the presence of Berry curvature but instead are a direct consequence of non-adiabatic 
injection of electrons from a contact
into the graphene. They are strongest for graphene with broken inversion symmetry
in the tunneling regime when the Fermi energy is in
the spectral energy gap around the Dirac point. 
Consequently in this regime these valley
currents are strongest close to the 
graphene/contact boundary. They are chiral and can be very strong close to the Dirac point, i.e., 
several times larger than
the conventional electric current even after averaging over the entire graphene nanostructure.
They can appear even at a voltage contact through which no {\em net} conventional electric
current flows provided that electrons are being emitted (and absorbed) by that contact.
They are predicted to be realized whenever electrons cross into the graphene at an 
abrupt boundary (which may
be regular or rough on the atomic scale) between
a contact and graphene with broken inversion symmetry at energies in the spectral gap around
the Dirac point. At Fermi energies well away from the Dirac point (i.e., 
outside of the gap in the density of states
of the nanostructure) valley currents can still be induced by bias voltages applied to the
nanostructure but in this regime they are considerably weaker, 
are not sensitive to whether or not the
inversion symmetry of the graphene is broken and are not confined to regions of the nanostructure
that are close to contacts. That valley currents can be induced in graphene nanostructures
even in the absence of inversion symmetry breaking has been recognized previously,\cite{Rycerz}
and is a consequence of the fact that the Bloch state wave vector need not be a conserved
quantity in nanostructures whose translational crystal symmetries are broken due to the presence
of boundaries.    

This research was supported by NSERC, CIFAR, Westgrid,
and Compute Canada.

{

\end{document}